\documentclass[12pt,english]{article}
\usepackage[T1]{fontenc}
\usepackage[latin9]{inputenc}
\usepackage{geometry}
\usepackage{amstext}
\usepackage{amsmath}
\usepackage{graphicx}
\usepackage{setspace}
\usepackage{esint}
\makeatletter

\newcommand{\be}{\begin{equation}}
\newcommand{\ee}{\end{equation}}

\newcommand{\noun}[1]{\textsc{#1}}

\newcommand{\lyxaddress}[1]{
\par {\raggedright #1
\vspace{1.4em}
\noindent\par}
}

\makeatother

\usepackage{babel}
\begin{document}

\vspace{-2cm}

\title{\noun{Functional and Local Renormalization Groups}}

\author{Alessandro Codello\emph{$^{\diamondsuit}$}, Giulio D'Odorico\emph{$^{\heartsuit}$} and Carlo Pagani\emph{$^{\clubsuit}$}}

\maketitle

\lyxaddress{\begin{center}
\emph{$^{\diamondsuit}$CP$^{\,3}$--Origins and Danish IAS,}\\
\emph{University of Southern Denmark,}\\
\emph{Campusvej 55, DK-5230 Odense M, Denmark}\\
\emph{}\\
\emph{$^{\heartsuit}$Radboud University Nijmegen,}\\
\emph{Institute for Mathematics, Astrophysics and Particle Physics,}\\
\emph{Heyendaalseweg 135, 6525 AJ Nijmegen, The Netherlands}\\
\emph{}\\
\emph{$^{\clubsuit}$ Institute of Physics, THEP,}\\
\emph{Johannes Gutenberg-University Mainz,}\\
\emph{Staudingerweg 7, 55099 Mainz, Germany}
\par\end{center}}

\begin{abstract}
We discuss the relation between functional renormalization group (FRG) and local
renormalization group (LRG), focussing on the two dimensional case as an example.
%
%Section 3
We show that away from criticality the Wess--Zumino action is described by a derivative expansion with coefficients naturally related to RG quantities.
%
%Section 4 and 5
We then demonstrate that the Weyl consistency conditions derived in the LRG approach are equivalent to the RG equation for the $c$--function
available in the FRG scheme. This allows us to give an explicit FRG representation of the Zamolodchikov--Osborn metric,
which in principle can be used for computations.\\
\\
Preprint: CP3-Origins-2015-003 DNRF90 and DIAS-2015-3
\end{abstract}
\tableofcontents{}

%%%%%%%%%%%
\section{Introduction}
%%%%%%%%%%%

The Renormalization Group (RG) is a key concept in Quantum Field Theory (QFT). 
%In standard QFT one usually starts with a free theory and perturbs it with weakly coupled interactions.
One usually starts by defining a QFT at some cutoff scale, after which the RG tells how the couplings
of the theory change when such scale is varied. In order for the theory to be well defined
up to arbitrarily high momenta the RG flow has to reach a fixed point as the cutoff is pushed to 
infinity. 
If we consider a fixed point action, a non trivial RG flow is triggered by breaking scale invariance, i.e. by adding relevant
or marginally relevant operators which start the flow. 
The ``directions'' of the breaking, defined by the beta functions of these operators,
can be seen in geometric terms as the initial velocities in theory
space, which is the manifold formed by the set of all couplings. 
This picture does not require a perturbative notion of renormalization to work. In fact,
in the Wilsonian approach, the ``bare'' or ``classical'' action
is simply the starting point of the flow, and the momentum shell integrations
to lower scales do not introduce infinities in the calculation. Infinities only appear if
we cannot choose a suitable initial condition for the flow such that
the cutoff scale can be removed, i.e. sent to infinity.

% Since the RG tells us how the dynamics changes when we change the
% scale, it can be understood as probing how the system responds under
% dilations. As we said, beta functions generate a nonzero divergence
% for the scale current, and break scale invariance. The Callan-Symanzik
% equation is in fact nothing more than the anomalous Ward identity
% for this broken scale invariance [Coleman]. If however one is interested in
% flows between two conformal points, or conformal field theories, one
% may seek an analogue equation for Weyl transformations (``local scale
% transformations''). Such equation was indeed studied by Osborn \cite{key-2,key-3,key-4}
% and is called the Local Renormalization Group (LRG) equation. It can
% be seen as the Ward identity for broken Weyl invariance, off-criticality,
% with the scale anomaly replaced by the trace anomaly.

% The standard machinery of weakly coupled QFTs cannot be straightforwardly
% applied to CFTs. interactions are there fully composite operators,
% eventually expressed in terms of primaries and descendants of the
% theory, and one needs to consider a source term for any of them in
% the generating functionals. This is one of the central features used
% in the LRG.

% Even after accounting for this fact though, only flows which originate
% from a CFT and remain in a sufficiently small neighborhood of it allow
% for a perturbative treatment of the deformation couplings. In the
% general case, conformal perturbation theory can only capture the initial
% phase of the flow, and if the IR CFT is sufficiently far, nonperturbative
% techniques are needed.

Having a complete understanding of the RG flow from the UV to the IR is in general a
very difficult task. Typically one considers a UV CFT and turns on a relevant operator
whose coupling can be treated perturbatively around the CFT. The entire flow in this case can be trusted 
only if the IR fixed point lies sufficiently close to the UV one. 
To understand the general case, thus, any constraint that can be put on the complete
flow can potentially give crucial information. In two dimensions such a constraint indeed exists, namely the so called $c-$theorem \cite{key-1-1},
which states that in every unitary Poincar{\'e} invariant theory there exists a function of the coupling
constants, the $c-$function, that decreases from UV to IR and that is stationary at the endpoints
of the flow, where its value equals the central charge of the corresponding CFT.
Nevertheless computing explicitly the $c-$function is a challenging problem. New insights
came thanks to the work of Osborn and collaborators \cite{key-2,key-3,key-4} who 
let the couplings be functions of spacetime and therefore act as sources for the composite
operators appearing in the action. Thanks to this it was possible to establish a more
direct path for the computation of the $c-$function and to understand in more detail 
the connection with the conformal anomaly. 
The relations found by Osborn come from imposing
the Wess--Zumino consistency conditions on the anomaly found in the theory where also 
the couplings are spacetime functions. Interestingly this approach is also successful in four dimension
where the $a-$theorem has been established \cite{key-6,key-7,key-8} and it hints at which may be the form of the $a-$function off-criticality.
Since the couplings are spacetime  dependent this approach is referred to as Local RG (LRG).

Another method which in principle allows to gain insight on all the RG flow
is given by the Functional, or exact, RG (FRG). In this approach one realizes the Wilsonian RG by
adding a suitable regulator term to the action. The modified generating functionals
defined in this way satisfy an exact RG equation. In particular, the use of a scale--dependent
effective action, the so--called effective average action,
offers many technical advantages \cite{key-14}.
The FRG stands out as a natural candidate to follow nonperturbatively
a complete RG trajectory. This was already investigated in \cite{key-1},
where a candidate $c-$function was constructed using the FRG.
It is then advisable to investigate
%how the Wess--Zumino consistency conditions constrain
%the FRG flows, how they can be implemented in FRG language, and what
%new information they uncover. This will also clarify
the connection
between FRG and LRG, and hopefully help establishing a common vocabulary
between the two.

% A complementary approach is to examine the global structure of theory
% space, to see what limitations are imposed on flows between different
% points of it. Symmetry and natural consistency requirements imposed
% on the theories that live in it can be sufficient to constrain the
% form of any possible (physically relevant) flow. The emblematic example
% of this is the c--theorem, in which imposing unitarity, Lorentz invariance
% and conservation of the energy momentum tensor one establishes the
% irreversibility of any RG flow, by showing that there always exists
% a function of the couplings that flows monotonically between fixed
% points \cite{key-1-1,key-6,key-7,key-8}.

% The $c$--theorem is strictly related to the LRG and the consistency
% conditions for Weyl symmetry. Indeed, one can impose an integrability
% condition on the trace anomaly, in the form of the Wess--Zumino consistency
% conditions (WZCC), which simply enforce the abelian nature of the
% Weyl group. The consistency conditions impose non--trivial relations
% between beta functions and couplings (local sources) in the LRG. While
% they constrain the form of the effective action at criticality, the
% major breakthrough comes from their application away from a fixed
% point. The c--theorem is then seen to be equivalent to the WZ consistency
% conditions applied to the anomaly off--criticality.

One could ask at this point why there should be a connection between
RG equations and Wess--Zumino consistency conditions. The physical idea that lies behind this is
the following. The RG can be thought of as a rescaling of the system; therefore, once we promote the RG scale to be spacetime dependent,
we are effectively considering local scale transformations of our
theory, i.e. Weyl transformations. The RG equation in terms of this
local scale, then, being connected to Weyl transformations, will be constrained
by the Wess--Zumino consistency conditions. The central result of the LRG is in fact
that the abelian character of the Weyl group can be used to obtain
statements about the RG flow in the form of consistency conditions.
In this work we will explore these issues for two dimensional theories.

The paper is organized as follows. We start in Section 2 by reviewing
the form of the effective action at fixed points, and the role of
the Wess--Zumino action. We then move away from criticality in Section
3. We will see that the form of the Wess--Zumino action off-criticality
can be constrained on very general grounds, and this allows for a
clear connection with the LRG. The latter will be exploited in Section
4, where we will also review how the Wess--Zumino consistency conditions leads to a derivation of
the $c-$theorem. In Section 5 instead we give an FRG representation
of the metric $\chi_{ij}$ introduced by Osborn.

%%%%%%%%%%%%%%
\section{Fixed point actions}
%%%%%%%%%%%%%%

Fixed point actions correspond to scale invariant theories. In two
dimensions
%, thanks to Zamolodchikov's theorem \cite{key-1-1},
we know that every fixed point theory represents a CFT. The problem in
studying CFT actions is that very few of them can be written in local
form. Notable examples are, apart from the Gaussian case, the fermionic
Ising model and the affine Kac--Moody actions. Here we will however maintain the discussion
on a general level, and will never need to resort to a specific
local form of the action.

From here on we will consider QFTs on a curved background characterized by a non--dynamical
metric $g_{\mu\nu}$. The motivation for working on a curved background
is not only one of generality, but one of convenience as well, since many things,
as for example the conformal anomaly to which now we turn, become
clearer and easier to describe in terms of curved space effective
actions.

\subsection{Conformal anomaly}

Consider a classically Weyl invariant theory defined in curved space.
Even if its classical energy momentum tensor is traceless, the diffeomorphism
invariant path integral measure used to quantize the theory is in
general not Weyl invariant and the quantum theory turns out to be
anomalous\footnote{See \cite{Mottola:1995sj} for more details.}:
\begin{equation}
\left\langle T_{\mu}^{\mu}\right\rangle =\frac{2}{\sqrt{g}}g_{\mu\nu}\frac{\delta\Gamma[g]}{\delta g_{\mu\nu}}\neq0\,.\label{CA_1}
\end{equation}
This is the trace, or conformal, anomaly; in $d=2$ its explicit form
is \cite{key-13}:
\begin{equation}
\left\langle T_{\;\mu}^{\mu}\right\rangle =-\frac{c}{24\pi}R\,,\label{CA_2}
\end{equation}
where $c$ is the conformal anomaly coefficient. In flat space this
expression formally vanishes; however, it gives contact terms in higher
order correlators. This is the way the anomaly manifests itself in
flat space. For instance, the two point function of the energy momentum
trace, in complex coordinates, turns out to be
\begin{equation}
\left\langle T_{zz}T_{ww}\right\rangle =\frac{1}{(2\pi)^{2}}\frac{c/2}{(z-w)^{4}}\,,\label{CA_3}
\end{equation}
which shows the equivalence of the anomaly coefficient with the central
charge of the theory.

\subsection{Wess--Zumino action}

While the conformal anomaly (\ref{CA_1}) represents the response
of the effective action to an infinitesimal Weyl transformation, the
response under a finite Weyl transformation is encoded in the Wess--Zumino
action, defined by:
\begin{equation}
\Gamma[e^{w\tau}\varphi,e^{2\tau}g]-\Gamma[\varphi,g]=c\,\Gamma^{WZ}[\tau,g]\,.\label{CA_4}
\end{equation}
We will refer to (\ref{CA_4}) as the Wess--Zumino relation. The linear
term in the Wess--Zumino action is engineered to give back the conformal
anomaly; in two dimensions it is possible to determine the full form
of the Wess--Zumino action by exploiting its relation with the Polyakov
action $S_{P}[g]=-\frac{1}{96\pi}\int\sqrt{g}R\frac{1}{\Delta}R$,
which is:
\begin{equation}
S_{P}[e^{2\tau}g]-S_{P}[g]=-\frac{1}{24\pi}\int d^{2}x\sqrt{g}\left[\tau\Delta\tau+\tau R\right]\equiv\Gamma^{WZ}[\tau,g]\,.\label{CA_5}
\end{equation}
Since the Weyl group is abelian, the Wess--Zumino action (\ref{CA_5})
is subject to a further constraint: the so called Wess--Zumino consistency
conditions. They essentially state that the order in which two successive
Weyl variations of the effective action are performed does not matter;
we will review them in Section 4.

\subsection{Fixed point effective action}

In this paper we will be interested in RG flows connecting two CFTs.
The considerations of the previous sections show that on a curved
background the effective action of any non--trivial CFT is not Weyl--invariant,
since any $c\neq0$ CFT is anomalous. Thus its fixed point effective
action must include a Polyakov term; in general it will have the following
split form \cite{key-1}:
\begin{equation}
\Gamma_{UV/IR}[\varphi,g]=S_{CFT_{UV/IR}}[\varphi,g]+c_{UV/IR}S_{P}[g]\,,\label{CA_6}
\end{equation}
at, respectively, the two endpoints of the flow $CFT_{UV}\,\to\, CFT_{IR}$
(if the IR theory has a mass gap, i.e. it is not a fixed point, then $c_{IR}=0$). Here $S_{CFT}$
is the curved space action for the CFT, which in flat space can be
defined via a Taylor expansion through its correlators, these being
in principle exactly known for any CFT. In this way the
Wess--Zumino relation (\ref{CA_4}) is trivially realized:
\begin{equation}
\Gamma[e^{w\tau}\varphi,e^{2\tau}g]-\Gamma[\varphi,g]=\underbrace{S_{CFT}[e^{w\tau}\varphi,e^{2\tau}g]-S_{CFT}[\varphi,g]}_{=0}+c\,(\underbrace{S_{R}[e^{2\tau}g]-S_{R}[g])}_{\Gamma^{WZ}[\tau,g]}\,.
\end{equation}
Having understood the general fixed point form of the effective action,
as in (\ref{CA_6}), we now turn to the problem of determining its
form away from criticality.

%%%%%%%%%%%%%%%
\section{Away from criticality}
%%%%%%%%%%%%%%%

We have seen how the form of the effective action is constrained at a critical point. 
After studying the fixed point structure of theory space, the next natural step is to consider
flows connecting different fixed points, which describe the cross--over from one critical point
to another. When we move away from a fixed point, the symmetry constraints
imposed on the action of course change: on one hand, scale invariance
is broken by the RG flow itself, and on the other hand new finite
terms can be generated by integrating the flow from the UV to the
IR. This expresses the fact that the RG breaking of scale invariance
adds further terms to the trace anomaly, and thus gives non--trivial
modifications to the Wess--Zumino action.

In this section we will investigate the form of the Wess--Zumino action
away from criticality. We will see that this is all we need to establish
a connection with the Local RG.% of Osborn et al.

\subsection{Running Wess--Zumino action}

When we flow away from a fixed point, the effective action will acquire
a scale dependence, which can be encoded in the so called effective average action, or just running effective action,
$\Gamma_{k}[\varphi,g]$, where $k$ is the scale.
If the perturbation which triggers the RG flow is composed of primary operators,
which transform homogeneously with respect to Weyl rescalings, it turns out that a generalized ``running''
Wess--Zumino action, defined by a scale dependent generalization of
the standard one, can be nicely constrained also away from criticality.
In the following we will always assume that the operators perturbing the CFT
are primaries.

The scale dependent, or running, Wess--Zumino action
$\Gamma_{k}^{WZ}[\tau,g]$
is defined by the following relation:
\begin{equation}
\Gamma_{ke^{-\tau}}[e^{w\tau}\varphi,e^{2\tau}g]-\Gamma_{k}[\varphi,g]=\Gamma_{k}^{WZ}[\tau,g]\,,
\label{AWC_1}
\end{equation}
which generalizes (\ref{CA_4}) away from criticality and reduces to it at any fixed point, where thus we must have:
%Thus $\Gamma_{k}^{WZ}[\tau,g]$ reduces to the standard Wess--Zumino
%action in this cases:
%
\begin{equation}
\Gamma_{UV/IR}^{WZ}[\tau,g]=c_{UV/IR}\Gamma_{WZ}[\tau,g]\,.\label{AWC_2}
\end{equation}
Equation (\ref{AWC_1}) is the starting point for all successive constructions of this paper.
The basic idea behind our construction is that it is much simpler to understand the structure of the running Wess--Zumino action
than that of the full running effective action. 
Note also that in (\ref{AWC_1}) we have rescaled $k$: this is the most natural choice since we are rescaling all dimensionful quantities. %, as is the scale $k$.
This choice makes the couplings appearing in the first running effective action on the lhs of (\ref{AWC_1}) implicitly spacetime dependent even if originally they were not;
as we will soon see, this fact will allow us to determine many properties of the running Wess--Zumino action as defined in (\ref{AWC_1}).
This way of thinking is similar to the one exposed in \cite{key-7} and \cite{key-10,key-12}, but different from the one employed in the LRG approach \cite{key-2,key-3}, that we will review in the next section, where couplings are taken to be space-time dependent from the beginning, i.e. they are treated as sources.  

We can start now to study the properties of the running Wess--Zumino action.
The first thing to notice is that there is no symmetry protecting the particular fixed point form (\ref{CA_4}), which will generically split away from criticality;
thus we may expect the following general form:
\begin{equation}
\Gamma_{k}^{WZ}[\tau,g]=-\frac{1}{24\pi}\int\sqrt{g}\left[\tilde{\mathcal{C}}_{k}\tau\Delta\tau+\mathcal{C}_{k}\tau R\right]+\beta\textrm{--terms}\,,
\label{AWC_3}
\end{equation}
with two, possibly different, running conformal anomaly coefficients $\tilde{\mathcal{C}}_{k}$ and $\mathcal{C}_{k}$.
Note that it is not clear at this point which running conformal anomaly coefficient will play the role of the $c$--function: this is the reason we used the calligraphic notation.
This problem becomes much more subtle in higher dimensions but we will not discuss it here.
Obviously terms that vanish at fixed points are also allowed, and will in general
be created by the RG flow: these are proportional to (dimensionless) beta functions\footnote{We will write explicitly the scale dependence of functions of the couplings $\beta^i, \omega_i,...$ since their dependence is implicit through that of the running couplings $g^i_k$.}
$\beta^{i}\equiv \partial_t g_k^{i}$ and are what we are calling $\beta$--terms.

The difference between the two running conformal anomalies that we
have introduced in (\ref{AWC_3}) must as well be proportional to
beta functions, $\mathcal{C}_{k}-\tilde{\mathcal{C}}_{k}=O(\beta)$,
since the two terms coincide at a fixed point.
Their difference will start with a term linear in the beta functions and can be written as
$O(\beta)=24\pi\omega_{i}\beta^{i}+...$, where $\omega_{i}$ is a form on the space of couplings.
The factor $24\pi$ is put just to connect with existing literature.
We may thus rewrite the running Wess--Zumino action as:
\begin{equation}
\Gamma_{k}^{WZ}[\tau,g]=-\frac{1}{24\pi}\int\sqrt{g}\left[\left(\mathcal{C}_{k}+24\pi\omega_{i}\beta^{i}\right)\tau\Delta\tau+\mathcal{C}_{k}\tau R\right]+\beta\textrm{--terms}\,.\label{AWC_3.1}
\end{equation}
%
%where obviously $\beta$--terms now stands for the remaining ones, which 
We will see in Section 4 that the combination,
\begin{equation}
c_{k}=\mathcal{C}_{k}+24\pi\omega_{i}\beta^{i}\,,\label{O_5}
\end{equation}
is indeed the correct one to be identified with the running $c$--function.
We turn now to better clarify the form of the other $\beta$--terms.

\subsection{Scale anomaly}

A clue about the form of the $\beta$--terms in (\ref{AWC_3}) comes
from the well known scale anomaly. If the perturbation away from criticality
is induced by some primary operators $\mathcal{O}_{i}$,
\begin{equation}
\Gamma_{k}[\varphi,g]=\Gamma_{UV}[\varphi,g]+\sum_{i}g^{i}_k\int\sqrt{g}\,\mathcal{O}_{i}[\varphi,g]\,,\label{SA_1}
\end{equation}
which define the couplings $g^{i}$ of dimension $[g^{i}]=k^{d_{i}}$,
then we know that the integral of the trace of the energy--momentum
tensor will have both a classical scale breaking piece, due to the
possible dimensionality of the couplings, and a quantum scale anomaly
proportional to the dimensionful beta function: 
\begin{equation}
\int\sqrt{g}\left\langle T_{\mu}^{\mu}\right\rangle _{k}=-\sum_{i}(\underbrace{\beta^{i}}_{\textrm{quantum}}-\underbrace{d_{i}g^{i}_k}_{\textrm{classical}})\int\sqrt{g}\,\mathcal{O}_{i}\,.\label{SA_2}
\end{equation}
Nicely enough the classical and quantum contributions combine to give
a term proportional to the dimensionless beta function:
\begin{equation}
\beta^{i}-d_{i}g^{i}_k=k^{d_{i}}\tilde{\beta}^{i}\,,\label{SA_3}
\end{equation}
where $\tilde{\beta}^{i}\equiv \partial_t \tilde{g}_k^{i}$ with $\tilde{g}_k^{i}=k^{-d_{i}}g_k^{i}$ the dimensionless couplings.
This is expected since this contribution to the trace anomaly is generated
by the RG flow, and therefore has to vanish at a fixed point, where
it is truly the dimensionless beta functions that vanish.

From the fact that the trace anomaly is essentially the variation
of the Wess--Zumino action with respect to the dilaton, we see that
the linear part of the $\beta$--terms must be of the following form:
\begin{equation}
\beta\textrm{--terms}=-k^{d_{i}}\tilde{\beta}^{i}\int\sqrt{g}\,\tau\,\mathcal{O}_{i}+O(\tau^{2}) \,.
\label{SA_4}
\end{equation}
%
%We can already suspect that the rhs of (\ref{SA_4}) will be a power
%series in $\tau\beta$, but we need some further ingredient before
%drawing this conclusion.
%
For simplicity, from now on we will consider only dimensionless couplings and thus drop the tilde in all subsequent formulas.

\subsection{Derivative expansion for the running Wess--Zumino action}

The information from the conformal and scale anomalies, encoded in
equations (\ref{AWC_3}) and (\ref{SA_4}), leads us to the natural
idea of considering a derivative expansion for the running Wess--Zumino action:
\begin{equation}
\Gamma_{k}^{WZ}[\tau,g]=\int\sqrt{g}\,\Big[V_{k}(\tau)+Z_{k}(\tau)\partial_{\mu}\tau\partial^{\mu}\tau+F_{k}(\tau)R\Big]+O(\partial^{4})\,,\label{DE_1}
\end{equation}
where from  (\ref{AWC_3}) and (\ref{SA_4}) we already know that:
\begin{eqnarray}
V_{k}(\tau) & = & -\tau\beta^{i}\mathcal{O}_{i}+...\nonumber \\
Z_{k}(\tau) & = & -\frac{\mathcal{C}_{k}}{24\pi}+\omega_{i}\beta^{i}+...\nonumber \\
F_{k}(\tau) & = & -\frac{\mathcal{C}_{k}}{24\pi}\tau+...\,.
\label{DE_2}
\end{eqnarray}
%
%since $\Gamma_{k}^{WZ}[0,g]=0$ implies that there are no constant terms.
For the moment let us also leave the order
of the next terms in (\ref{DE_2}) unspecified, the reason will become
clear in a second.
Note that the $c$--function, and thus (modulo beta functions) also the running anomaly coefficients in (\ref{DE_2}), are related to the beta function of Newton's gravitational constant \cite{key-1,Codello:2014wfa} and so the three functions in (\ref{DE_2}) are of the same linear order in the beta functions.

An important fact is that matter fields $\varphi$ enter only in
the potential term, i.e. only $V_k$ and not $Z_k$ or $F_k$ depend on $\varphi$.
This is a consequence of the fact that we are considering primary perturbations of the fixed point action,
for which mixed derivative terms of the form $\partial_{\mu}\varphi\partial^{\mu}\tau$
(or more general) are not created in the difference between the effective actions in (\ref{AWC_1}).
This is the main reason to consider primary perturbations in (\ref{SA_1});
more general deformations can still in principle be treated, at the price of losing simplicity.

A shortcut to find out the form of the higher order corrections to
the potential term comes from exploiting the fact that our definition of the running
Wess--Zumino action (\ref{AWC_1}) contains the rescaling $k\rightarrow e^{-\tau}k$
%of the scale $k$ by a factor
which renders the couplings formally spacetime dependent
%
%\begin{equation}
%$k\rightarrow e^{-\tau}k$
%\qquad\qquad
$g_{k}^{i}\rightarrow g_{ke^{-\tau}}^{i}$. %\,.\label{ST_1}
%\end{equation}
%
As we said, the main difference with respect to the LRG \cite{key-2,key-3} is that we promote the couplings
to be spacetime dependent in a particular manner, namely via the
%transformation $k\rightarrow ke^{-\tau}$
rescaling of $k$; similarly to what has been considered in \cite{key-7,key-8}.
Expanding now the rescaled couplings in powers of $\tau$ we get:
\begin{eqnarray}
g_{ke^{-\tau}}^{i} & = & g_{k(1-\tau+...)}^{i}=g_{k}^{i}-\tau\beta^{i}+O(\tau^{2})\,.\label{ST_2}
\end{eqnarray}
If we use (\ref{ST_2}) in the primary deformation introduced in equation
(\ref{SA_1}) and insert in the off--critical Wess--Zumino relation (\ref{AWC_1}) we immediately recover the scale anomaly part of the $\beta$--terms:
\begin{eqnarray}
\Gamma_{e^{-\tau}k}[e^{w\tau}\varphi,e^{2\tau}g]-\Gamma_{k}[\varphi,g] & = & \int\sqrt{g}\left[(g_{k}^{i}-\tau\beta^{i})\mathcal{O}_{i}-g_{k}^{i}\mathcal{O}_{i}\right]+O(\tau^{2})\nonumber \\
 & = & -\int\sqrt{g}\,\tau\beta^{i}\mathcal{O}_{i}+O(\tau^{2})\,.\label{ST_3}
\end{eqnarray}
But now we can also look at the higher order terms appearing in the expansion of the couplings,
\begin{eqnarray}
g_{ke^{-\tau}}^{i}=g_{k}^{i}-\tau\beta^{i}+\frac{1}{2}\tau^{2}\beta^{j}\partial_{j}\beta^{i}+O(\tau^{3})\,,
\end{eqnarray}
which, when used in (\ref{ST_3}), lead us to the following intriguing expansion for the potential \cite{key-8}:
\begin{equation}
V_{k}(\tau)=\left[-\beta^{i}\tau+\frac{1}{2}\beta^{j}\partial_{j}\beta^{i}\tau^{2}\right]\mathcal{O}_{i}+O(\tau^{3})\,.\label{ST_4}
\end{equation}
The same reasoning can be applied to the other functions $Z_{k}$ and $F_{k}$.
For instance, since the running anomaly coefficients are the couplings of the Polyakov action, they can be treated as in (\ref{ST_2}):
\begin{equation}
\mathcal{C}_{ke^{-\tau}}=\mathcal{C}_{k}-\tau\partial_{t}\mathcal{C}_{k}+O(\tau^{2})
\end{equation}
and similarly for $\omega_{i}\beta^{i}$ since they are also functions of the couplings\footnote{More precisely, they are the coupling of a non--local term obtained by eliminating the dilaton, as described shortly.}.
Thus the $\beta$--terms, including the $\omega_i \beta^i$ piece, become:%\footnote{We are dismissing terms with mixed internal--spacetime indices.}:
\begin{eqnarray}
\beta\textrm{--terms} & = & \int\sqrt{g}\left\{ \Big[-\tau\beta^{i}+\frac{1}{2}\tau^{2}\beta^{j}\partial_{j}\beta^{i}+...\Big]\mathcal{O}_{i}\right.\nonumber \\
% &  & +\left(-\omega_{i}\beta^{i}\right)\partial_{\mu}\tau\partial^{\mu}\tau\nonumber \\
 &  & \left.+\left(-\omega_{i}\beta^{i}\right)\partial_{\mu}\tau\partial^{\mu}\tau+\left[\partial_{t}\left(\omega_{i}\beta^{i}\right) + ...
% -\chi_{ij}\beta^{i}\beta^{j}
 \right]\tau\partial_{\mu}\tau\partial^{\mu}\tau\right\} +O(\tau^{4})\,.\label{O_3}
\end{eqnarray}
%
%We have here introduced another coupling dependent term of second
%order in the beta functions, constructed naturally with the ``metric''
%$\chi_{ij}$, to complete the $\tau\partial_{\mu}\tau\partial^{\mu}\tau$
%terms. Notice that according to the discussion so far, in the second
%and third lines there can in principle appear further contributions
%of higher beta order, however these simply ammount to a redefinition
%of $\chi_{ij}$ and $\omega_{i}$, which are in any case unspecified
%functions of the couplings. The previous expansion can thus also be
%seen as implicitly defining $\chi_{ij}$ and $\omega_{i}$.
%
Putting all terms together we finally arrive at following form for
the derivative expansion of the running Wess--Zumino action:
\begin{eqnarray}
V_{k}(\tau) & = & \left[-\beta^{i}\tau+\frac{1}{2}\beta^{j}\partial_{j}\beta^{i}\tau^{2}\right]\mathcal{O}_{i}+O(\tau^{3})\nonumber \\
Z_{k}(\tau) & = & -\frac{\mathcal{C}_{k}}{24\pi}-\omega_{i}\beta^{i}+\left[\partial_{t}\left(\frac{\mathcal{C}_{k}}{24\pi}+\omega_{i}\beta^{i}\right) + ...
%-\chi_{ij}\beta^{i}\beta^{j}
\right]\tau+O(\tau^{2})\nonumber \\
F_{k}(\tau) & = & -\frac{\mathcal{C}_{k}}{24\pi}\tau+\left[\partial_{t}\left(\frac{\mathcal{C}_{k}}{24\pi}\right)+...\right]\tau^{2}+O(\tau^{3})\,.
\label{O_4}
\end{eqnarray}
The dots stand for additional terms that are not scale derivatives of lower order terms
and thus cannot be derived by previous reasoning; even if in principle one could make an ansatz at this point, we will not discuss them now since the terms relevant to our discussion in Section 5 are all already present.
%will wait the next section to discuss these terms.
%
%We notice that the combination,
%%
%\begin{equation}
%c_{k}=\mathcal{C}_{k}+24\pi\omega_{i}\beta^{i}\,,\label{O_5}
%\end{equation}
%%
%naturally appears in the expansion of $Z_k$; we will soon see that indeed this combination is the correct one to be identified with the running $c$--function.
%
%We will in fact see in the next section that, using the Wess--Zumino
%consistency conditions in the way they are used in the LRG, the
%coefficient of the $\tau\partial_{\mu}\tau\partial^{\mu}\tau$ term
%in (\ref{O_4}) has to vanish.
%
%{\it This requirement, combined with unitarity, gives the $c$--theorem.}
%
%However, from here one could also directly proceed to study the RG
%flow of the running Wess--Zumino action based on the ansatz (\ref{O_4}). Indeed,
%if one wants to use the $c$--theorem, for instance to compute the
%central charges of interacting theories, the general analysis presented
%up to now is empty until we have a consistent procedure to compute
%the beta functions together with the flow of $c_{k}$. This is what
%we will describe in Section 5, using the FRG.

Finally, as a small aside let us make the following remark. The advantage of
working with the running Wess--Zumino action is that being it written
in terms of $\tau$ it is local and thus expandible in a derivative
expansion. However, formally, we can get rid of the dilaton at any
point of the previous construction, by considering the function $\tau(g)$,
taken as a solution to the ``equation of motion'' $\Delta\tau=R$.
This construction is useful for different purposes, as was discussed in \cite{key-1}.
For example it can be used to eliminate $\tau$ from the off--critical Wess--Zumino relation (\ref{AWC_1}),
in this way leading to the explicit form of the running effective action away from criticality.
Notice also that the on--shell condition $\tau=\tau(g)$
reduces in flat space to the requirement that the dilaton satisfies
$\partial^2 \tau=0$, which is the condition used in \cite{key-5}.

%%%%%%%%%%%%%%%%%%%%
\section{Local Renormalization Group}
%%%%%%%%%%%%%%%%%%%%

%In the previous section we focused our attention on the Wess--Zumino action
%away from a fixed point. This led us to consider a derivative expansion
%for the dilaton and to add some new terms proportional to the beta
%functions, which we called $\beta$--terms. This expansion already
%encodes non--trivial information about the general form of the effective
%action away from criticality, but is not enough to fully understand the $\beta$--terms.
%
In this section we review, and contextualize along the lines of the previous sections,
%Further insight can be gained considering
the LRG approach first proposed
by Osborn and collaborators in a series of works \cite{key-2,key-3} and recently resumed in \cite{key-4,key-5,Grinstein:2013cka}.
The LRG approach is based on the idea of promoting the couplings to fields $g^{i}\to g^{i}(x)$
so that they play the role of sources for their corresponding operators $\mathcal{O}_{i}$. 
Furthermore the couplings, being now explicitly spacetime dependent functions, are
responsible for new terms in the conformal anomaly (\ref{CA_2}).
The latter, when combined with the Wess--Zumino consistency conditions, lead to
extremely useful relations between beta functions and other RG quantities like the $c$--function.
%
%
%{\it In the language of the previous section, these relations appear as constraints on the coefficients of the
%derivative expansion of the running Wess--Zumino action.}

\subsection{Osborn's ansatz}

%In this section we consider in detail the relation between the form of the
%Wess--Zumino action proposed in the previous section and the LRG approach.
%
As a first step we consider the ansatz made by Osborn for the conformal anomaly. 
Since the couplings in the LRG approach are explicitly spacetime dependent new
terms will appear in the conformal anomaly away from criticality. To linear order in
the parameter of the transformation the LRG running Wess--Zumino action reads:
\begin{equation}
%\int\sqrt{g}\left\langle T_{\mu}^{\mu}\right\rangle+
\Gamma_{k}^{WZ}[\tau,g]=\int\sqrt{g}\left[-\tau\beta^{i}\mathcal{O}_{i}+\chi_{ij}\partial_{\mu}g^{i}_k\partial^{\mu}g^{j}_k\tau+\omega_{i}\partial_{\mu}\tau\partial^{\mu}g^{i}_k-\frac{\mathcal{C}_{k}}{24\pi}\tau R\right]+O(\tau^2)\,,\label{O_1}
\end{equation}
%+O(\tau^{2})
%
where $\chi_{ij}$ and $\omega_{i}$ are arbitrary functions
of the couplings  while $\mathcal{C}_{k}$ is the running anomaly coefficient
($\omega_{i}$ and $\mathcal{C}_{k}$ are in principle different from the ones defined in the previous section).
A further term proportional to a current $J^\mu_i$ has been considered in \cite{Friedan:2009ik}
but we will neglect such term in our discussion.
Equation (\ref{O_1}) is the starting point of the LRG analysis, which
then uses the Wess--Zumino consistency conditions to derive non--trivial
relations between $\mathcal{C}_{k}$, $\omega_{i},$
$\chi_{ij}$ and the beta functions $\beta^{i}$.

\subsection{Wess--Zumino consistency conditions}

The abelian nature of the Weyl group implies that the fixed point
Wess--Zumino action should satisfy the Wess--Zumino consistency condition \cite{key-17}:
\begin{equation}
\Gamma^{WZ}[\tau_{1},e^{2\tau_{2}}g]-\Gamma^{WZ}[\tau_{1},g]=\Gamma^{WZ}[\tau_{2},e^{2\tau_{1}}g]-\Gamma^{WZ}[\tau_{2},g]\,,\label{WZ_1}
\end{equation}
which simply states that any two finite Weyl variations of the fixed
point action must commute. We may now expand:
\begin{equation}
\Gamma^{WZ}[\tau_{1},e^{2\tau_{2}}g]=\Gamma^{WZ}[\tau_{1},g]+\delta_{\tau_{2}}\Gamma^{WZ}[\tau_{1},g]+...
\end{equation}
where $\delta_{\tau}\equiv\int2\tau g_{\mu\nu}\frac{\delta}{\delta g_{\mu\nu}}$
and obtain the infinitesimal version of (\ref{WZ_1}): 
\begin{equation}
\delta_{\tau_{2}}\Gamma^{WZ}[\tau_{1},g]=\delta_{\tau_{1}}\Gamma^{WZ}[\tau_{2},g]\,.\label{WZ_2}
\end{equation}
One can check explicitly that the fixed point Wess--Zumino action
(\ref{CA_5}) satisfies (\ref{WZ_1}) or (\ref{WZ_2}).

The Wess--Zumino consistency conditions are also valid away from criticality
(since they encode a property of the Weyl group that has nothing to
do with the fixed point) and can be imposed on the running Wess--Zumino
action defined in equation (\ref{AWC_1}):
%
%\begin{eqnarray}
%\Gamma_{ke^{-\tau_{2}}}^{WZ}[\tau_{1},e^{2\tau_{2}}g]-\Gamma_{k}^{WZ}[\tau_{1},g] & = & \left(\Gamma_{ke^{-\tau_{2}}e^{-\tau_{1}}}[e^{2\tau_{1}}e^{2\tau_{2}}g]-\Gamma_{ke^{-\tau_{2}}}[e^{2\tau_{2}}g]\right)-\left(\Gamma_{ke^{-\tau_{1}}}[e^{2\tau_{1}}g]-\Gamma_{k}[g]\right)\nonumber \\
% & = & \left(\Gamma_{ke^{-\tau_{1}}e^{-\tau_{2}}}[e^{2\tau_{2}}e^{2\tau_{1}}g]-\Gamma_{ke^{-\tau_{1}}}[e^{2\tau_{1}}g]\right)-\left(\Gamma_{ke^{-\tau_{2}}}[e^{2\tau_{2}}g]-\Gamma_{k}[g]\right)\nonumber \\
% & = & \Gamma_{ke^{-\tau_{1}}}^{WZ}[\tau_{2},e^{2\tau_{1}}g]-\Gamma_{k}^{WZ}[\tau_{2},g]\,,\label{WZ_2.1}
%\end{eqnarray}
%
\begin{align}
\Gamma_{ke^{-\tau_{2}}}^{WZ}[\tau_{1},e^{2\tau_{2}}g]-\Gamma_{k}^{WZ}[\tau_{1},g] &=  \left(\Gamma_{ke^{-\tau_{2}}e^{-\tau_{1}}}[e^{2\tau_{1}}e^{2\tau_{2}}g]-\Gamma_{ke^{-\tau_{2}}}[e^{2\tau_{2}}g]\right)-\left(\Gamma_{ke^{-\tau_{1}}}[e^{2\tau_{1}}g]-\Gamma_{k}[g]\right)\nonumber \\
%  \end{equation}
 %\begin{equation}
& =  \left(\Gamma_{ke^{-\tau_{1}}e^{-\tau_{2}}}[e^{2\tau_{2}}e^{2\tau_{1}}g]-\Gamma_{ke^{-\tau_{1}}}[e^{2\tau_{1}}g]\right)-\left(\Gamma_{ke^{-\tau_{2}}}[e^{2\tau_{2}}g]-\Gamma_{k}[g]\right)\nonumber \\  \label{WZ_2.1}
% \end{equation}
 &=  \Gamma_{ke^{-\tau_{1}}}^{WZ}[\tau_{2},e^{2\tau_{1}}g]-\Gamma_{k}^{WZ}[\tau_{2},g]\,.%\qquad\qquad\qquad\qquad\qquad\qquad\qquad\,
\end{align}
The abelian character of the Weyl transformations has been used in second line when we
exchanged the order of appearance of $\tau_{1}$ and $\tau_{2}$. To linear order we can write:
\begin{eqnarray}
\Gamma_{ke^{-\tau}}^{WZ}[\sigma,e^{2\tau}g] & =\Gamma_{k}^{WZ}[\sigma,g]+  \Delta_{\tau}\Gamma_{k}^{WZ}[\sigma,g]+O\left(\tau^{2}\right)\,,\label{WZ_2.2}
\end{eqnarray}
defining the operator implementing off--critical infinitesimal Weyl transformations:
\begin{equation}
\Delta_{\tau}\equiv\int d^{2}x\,\tau\left\{ 2g_{\mu\nu}\frac{\delta}{\delta g_{\mu\nu}}-\beta^{i}\frac{\delta}{\delta g^{i}}\right\} \,.\label{WZ_2.3}
\end{equation}
Using (\ref{WZ_2.2}) in both sides of (\ref{WZ_2.1}) leads to the
infinitesimal Wess--Zumino consistency conditions:
\begin{equation}
\Delta_{\tau_{2}}\Gamma_{k}^{WZ}[\tau_{1},g]=\Delta_{\tau_{1}}\Gamma_{k}^{WZ}[\tau_{2},g]\,.\label{WZ_CC}
\end{equation}
Inserting the ansatz (\ref{O_1}) in the Wess--Zumino consistency conditions (\ref{WZ_CC}) leads to different useful relations \cite{key-2,key-3},
which ultimately lead to the fundamental two dimensional (Weyl) consistency condition:
%and usethe expansion (\ref{O_2}) we find:
%
%\begin{eqnarray}
%%\Gamma_{ke^{-\tau_{2}}}^{WZ}[\tau_{1},e^{2\tau_{2}}g]-\Gamma_{k}^{WZ}[\tau_{1},g] & = & \int d^{2}x\sqrt{g}\left\{ \frac{1}{2}\tau_{1}\tau_{2}\beta^{j}\partial_{j}\beta^{i}\mathcal{O}_{i}+\left(-\omega_{i}\beta^{i}\right)\partial_{\mu}\tau_{1}\partial^{\mu}\tau_{2}\right.\nonumber \\
% &  & \left.+\left[\partial_{t}\left(\frac{\mathcal{C}_{k}}{24\pi}+\omega_{i}\beta^{i}\right)-\chi_{ij}\beta^{i}\beta^{j}\right]\tau_{1}\partial_{\mu}\tau_{2}\partial^{\mu}\tau_{2}\right\} +...\label{WZ_CC_1}
%\end{eqnarray}
%%
%When acting with the operator (\ref{WZ_2.3}) on (\ref{WZ_CC_1})
%we effectively have:
%%
%\begin{eqnarray*}
%\Delta_{\tau} & = & \int d^{2}x\,\tau\left\{ \frac{\delta g_{\mu\nu}}{\delta\tau}\frac{\delta}{\delta g_{\mu\nu}}+\frac{\delta g^{i}}{\delta\tau}\frac{\delta}{\delta g^{i}}\right\} =\tau\frac{\delta}{\delta\tau}\,,
%\end{eqnarray*}
%%
%and we see that the coefficient of $\tau_{1}\partial_{\mu}\tau_{2}\partial^{\mu}\tau_{2}$
%must vanish:
%
\begin{equation}
\partial_{t}\left(\frac{\mathcal{C}_{k}}{24\pi}+\omega_{i}\beta^{i}\right)=\chi_{ij}\beta^{i}\beta^{j}\,.\label{WZ_CC_2}
\end{equation}
%
%This is the the two dimensional Weyl consistency condition found by Osborn \cite{key-3}.
After identifying the real $c$--function as did in (\ref{O_5}) we find:
\begin{equation}
\partial_{t}c_{k}=24\pi\chi_{ij}\beta^{i}\beta^{j}\,,
\label{CC_1}
\end{equation}
which shows that $\chi_{ij}$ plays the role of Zamolodchikov's metric.
Reflection--positivity allows to prove that Zamolodchikov's metric is positive definite,
thus implying the $c$--theorem for theories having this property  \cite{key-1-1,key-3} .

But in general, these relations are empty (we can say that they are only kinematical)
until all the objects entering in equation (\ref{CC_1}) have been explicitly defined or constructed:
we need a way to compute both the beta functions and the metric $\chi_{ij}$.
One way to achieve this is to use perturbation theory or conformal perturbation theory \cite{key-2,key-3}.
Another possibility is to use the exact RG equations as we will do in the next section. 
% Note
% that in any case, the Weyl consistency conditions give relations between
% beta functions, but they don't tell how to compute them and one must
% rely on some technique, such as perturbation theory or conformal perturbation
% theory.
%
%{\it As said previously, this tells us something about the derivative
%expansions of the running WZ action: the linear term in the expansion
%of $Z_{k}(\tau)$ in equation (\ref{O_4}) vanishes if the suspension dots are to be replaced by $-\chi_{ij}\beta^{i}\beta^{j}$.}

%%%%%%%%%%%%%%%%%%%%%%
\section{Functional Renormalization Group}
%%%%%%%%%%%%%%%%%%%%%%

\subsection{Flow equation for $c_{k}$}

The general form of the running Wess--Zumino action has no content until we
choose a regularization scheme to compute the beta functions and the
RG flow. In the LRG, it is usually implicitly assumed a standard scheme
such as dimensional regularization; here we will use the non--perturbative
FRG scheme, as a continuation of the analysis given in \cite{key-1}.

In the FRG the scale dependence of the effective average action
functional $\Gamma_{k}[\varphi,g]$ is governed by an exact equation \cite{key-15}:
\begin{equation}
\partial_{t}\Gamma_{k}[\varphi,g]=\frac{1}{2}\textrm{Tr}\left(\frac{\delta^{2}}{\delta\varphi\delta\varphi}\Gamma_{k}[\varphi,g]+R_{k}[g]\right)^{-1}\partial_{t}R_{k}[g]\,.\label{RG_3}
\end{equation}
In order to solve this equation one has to choose a truncation for
the running effective action, and a suitable regulator term $R_{k}$. For a local
truncation of the form (\ref{SA_1}), for instance, one can extract
all beta functions by expanding both sides of equation (\ref{RG_3})
in the $\int\sqrt{g}\,\mathcal{O}_{i}$ operator basis:
\begin{eqnarray}
\partial_{t}\Gamma_{k}[\varphi,g] & = & \sum_{i}\partial_{t}g^{i}_k\int \sqrt{g}\,\mathcal{O}_{i}[\varphi,g]\nonumber\\
\frac{1}{2}\textrm{Tr}\left(\frac{\delta^{2}}{\delta\varphi\delta\varphi}\Gamma_{k}[\varphi,g]+R_{k}\right)^{-1}\partial_{t}R_{k} & = & \sum_{i}\beta^{i}\int \sqrt{g}\,\mathcal{O}_{i}[\varphi,g]\,.
\label{RG_3.1}
\end{eqnarray}
Comparing the two results one can read off the beta functions, in principle without the use of any perturbative expansion.

This logic can now be applied to the $c$--function as well to compute
its running. From the discussion in Section 3 one is naturally led
to identify the running $c$--function as the coefficient of the term
$\int\sqrt{g}(\partial\tau)^{2}$ in the running Wess--Zumino action (see equations (\ref{O_4}) and (\ref{O_5})). The flow
equation for the running Wess--Zumino action can be easily found by taking a scale derivative of its definition (\ref{AWC_1}):
\begin{equation}
\partial_{t}\Gamma_{k}^{WZ}[\tau,g]=\partial_{t}\Gamma_{e^{-\tau}k}[e^{w\tau}\varphi,e^{2\tau}g]-\partial_{t}\Gamma_{k}[\varphi,g]\,.\label{RG_1}
\end{equation}
By projecting out the term proportional to $\int\sqrt{g}(\partial\tau)^{2}$
we immediately derive the RG equation for the $c$--function:
\begin{eqnarray}
\partial_{t}c_{k} & = & -24\pi\left.\partial_{t}\Gamma_{e^{-\tau}k}[e^{w\tau}\varphi,e^{2\tau}g]\right|_{\int\sqrt{g}(\partial\tau)^{2}}\nonumber \\
 & = & -12\pi\left.\textrm{Tr}\left(\frac{\delta^{2}}{\delta\varphi\delta\varphi}\Gamma_{e^{-\tau}k}[e^{w\tau}\varphi,e^{2\tau}g]+R_{e^{-\tau}k}[e^{2\tau}g]\right)^{-1}\partial_{t}R_{e^{-\tau}k}[e^{2\tau}g]\right|_{\int\sqrt{g}(\partial\tau)^{2}}\,.\label{RG_2}
\end{eqnarray}
%
%Note that the flow of the $c$--function is determined by the flow
%of the two--point function of the dilaton.
%
The cutoff action is Weyl invariant when $k$ is rescaled as in (\ref{RG_2}); we can thus set the metric to be the flat
one $g_{\mu\nu}=\delta_{\mu\nu}$ and use $R_{e^{-\tau}k}[e^{2\tau}g]=R_{k}[g]$ to write:
\begin{equation}
\partial_{t}c_{k}=-12\pi\left.\textrm{Tr}\left(\frac{\delta^{2}}{\delta\varphi\delta\varphi}\Gamma_{k}[e^{w\tau}\varphi,e^{2\tau}\delta]+R_{k}[\delta]\right)^{-1}\partial_{t}R_{k}[\delta]\right|_{\int(\partial\tau)^{2}}\,,\label{RG_5}
\end{equation}
The explicit $\int(\partial\tau)^{2}$ terms on the rhs of (\ref{RG_5}) can be selected by first using equation (\ref{AWC_1}) to make explicit the dilaton dependence; then by taking two dilaton functional derivatives of the trace and  
% of the trace and then setting  with respect to the dilaton This equation shows that the flow equation for the dilaton two point function can be derived
%Finally, after taking two functional derivatives with respect to the dilatonof the flow equation for the
%effective average action (\ref{RG_3}); in order to extract the running of the $c$--function we just
setting $\varphi=\tau=0$; and finally by picking the order $p^{2}$ terms:
\begin{figure}
\begin{centering}
\includegraphics[scale=0.4]{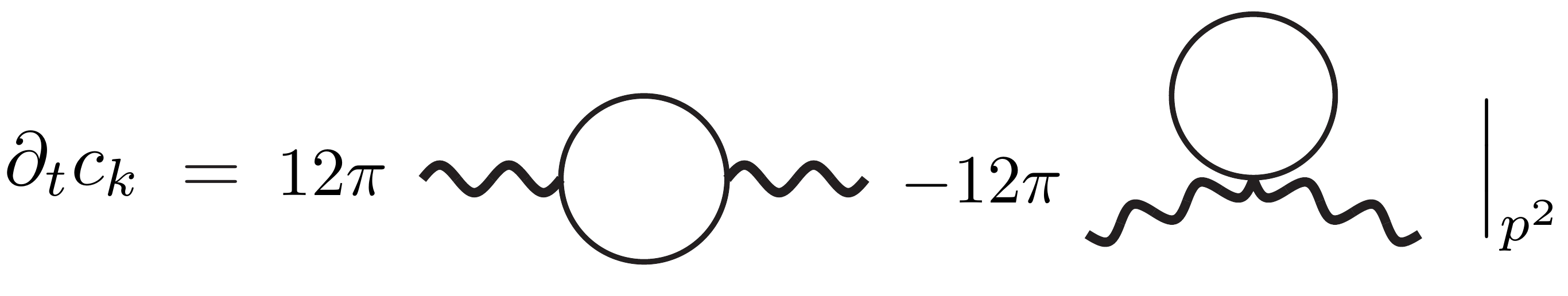}
\caption{Diagrammatic representation of the flow equation of the $c$--function as given in equation (\ref{RG_5.1}).}
%\par
\end{centering}
\end{figure}
%%
%\begin{eqnarray}
%\partial_{t}c_{k} & = & \left.-12\pi\mbox{Tr}\left[G_{k}\frac{\delta^{3}\Gamma_{k}^{WZ}}{\delta\tau\delta\varphi\delta\varphi}G_{k}\frac{\delta^{3}\Gamma_{k}^{WZ}}{\delta\tau\delta\varphi\delta\varphi}G_{k}\partial_{t}R_{k}\right]\right|_{p^{2}}\nonumber \\
% &  & +6\pi\left.\mbox{Tr}\left[G_{k}\frac{\delta^{4}\Gamma_{k}^{WZ}}{\delta\tau\delta\tau\delta\varphi\delta\varphi}G_{k}\partial_{t}R_{k}\right]\right|_{p^{2}}\,,\label{RG_5.1}
%\end{eqnarray}
%%
%
\begin{eqnarray}
\partial_{t}c_{k}  =  \left.12\pi\,\mbox{Tr}\,\tilde{\partial}_t\left\{G_{k}\frac{\delta^{3}\Gamma_{k}^{WZ}}{\delta\tau\delta\varphi\delta\varphi}G_{k}\frac{\delta^{3}\Gamma_{k}^{WZ}}{\delta\tau\delta\varphi\delta\varphi}\right\}\right|_{p^{2}}-12\pi\left.\mbox{Tr}\,\tilde{\partial}_t\left\{G_{k}\frac{\delta^{4}\Gamma_{k}^{WZ}}{\delta\tau\delta\tau\delta\varphi\delta\varphi}\right\}\right|_{p^{2}}\,,\label{RG_5.1}
\end{eqnarray}
where $G_{k}\equiv\left(\Gamma_{k}^{\left(2,0\right)}+R_{k}\right)^{-1}$
and $\tilde{\partial}_t\equiv \partial_t R_k \frac{\partial}{\partial R_k}$ (for more details see  \cite{key-1}).
%Obviously only the running Wess--Zumino action enters the equation for the $c$--function.
This is the explicit flow equation for the $c$--function and it can be represented as in Figure 1.
The non--trivial running is due to the interaction vertices between the matter fields
(which run in the loop) and the dilaton\footnote{The reader may wonder why there are no dilaton vertices acting on
the cutoff kernel. Let us note that the rescaling of the cutoff, combined
with the Weyl rescaling of the matter fields, allows to avoid $\tau$
dependences in the cutoff action $\Delta S_{k}$. This is very convenient
for two reasons: first we avoid some scheme dependent contributions
(i.e. dependent on the specific form of the cutoff action). Second
this simplifies the computation since there are no dilaton vertices
coming from $\Delta S_{k}$. Note that actually $\Delta S_{k}$ is
invariant under simultaneous variation of the fields and the cutoff
only if the operator used in the cutoff kernel is Weyl covariant.
This will always be the case in our computations.}.
%
%It is clear that the interesting terms of (\ref{RG_5.1}) are those up to order $\tau^{2}$.
%In particular the terms up to order $\tau^{2}$ which interact with the matter fields
%are the most relevant since are those which contribute to the running
%of the $c$--function.

\subsection{Recovering the consistency conditions}

We are now ready to combine the flow equation for the $c$--function
(\ref{RG_5.1}) with the form of the derivative expansion of the running
Wess--Zumino action we found in Section 3.

Since the vertices in the two diagrams of Figure 1 are matter--dilaton vertices, only the potential will contribute.
This is a very important fact and will lead to a very simple form for the flow of the $c$--function.
The first diagram of Figure 1, representing the first trace in equation (\ref{RG_5.1}),
will involve the three point vertex stemming from the monomial $\tau\beta^{i}{\cal O}_{i}$,
%:
%%
%\begin{equation}
%\frac{\delta^{3}\Gamma_{k}^{WZ}}{\delta\tau\delta\varphi\delta\varphi}
%%=\Gamma_{k}^{\left(2,1\right)}\left[\varphi,\tau\right] & \supseteq & \frac{\delta^{3}}{\delta\tau\delta\varphi\delta\varphi}\left[\int\tau\beta_{i}{\cal O}_{i}\right]
%=\beta_{i}\frac{\delta^{2}{\cal O}_{i}}{\delta\varphi\delta\varphi}\,,
%\label{RG_6.01}
%\end{equation}
%
since all other higher order terms in the beta functions will not contribute once the dilaton is set to zero in the vertex.
In this diagram the $p^2$ dependence is entirely due to one of the $G_k$ which is evaluated at $(q+p)^2$.
The second diagram, representing the second trace in equation (\ref{RG_5.1}),
will instead involve a vertex derived from $\tau^{2}\beta_{j}\partial_{j}\beta_{i}{\cal O}_{i}$.
This however does not contribute to the running of the $c$--function
since there are no derivatives acting on the dilaton producing $p^2$ contributions (remember that
our deformations are primaries),
and these cannot come from the $G_k$ since they are evaluated at $q^2$.

Thus we conclude that only the first diagram contributes: this implies that
the final form of the flow equation is quadratic in the (dimensionless) beta functions.
More precisely we find:
\begin{equation}
\partial_{t}c_{k}=24\pi\chi_{ij}\beta^{i}\beta^{j}\,,
\label{RG_6}
\end{equation}
where the FRG's explicit form for Zamolodchikov's metric is:
\begin{equation}
\chi_{ij}=\frac{1}{24\pi}\int\frac{d^{2}q}{(2\pi)^{2}}\tilde{\partial}_{t}\left\{ G_{k}(q^{2})G_{k}\left((q+p)^{2}\right)\right\} \mathcal{O}_{i}^{(2)}(q,q+p)\mathcal{O}_{j}^{(2)}(-q-p,-q)\,.
\label{Zam}
\end{equation}
Here $\mathcal{O}_{i}^{(2)}(q_{1},q_{2})$ denotes the momentum space representation of the vertex $\frac{\delta^2 \mathcal{O}_i}{\delta \varphi \delta \varphi}$.
These last two relations are the main result of this paper.
It is clear that the FRG flow equation for the $c$--function (\ref{RG_6}) is exactly the same
as the Weyl consistency condition (\ref{CC_1}) derived within the LRG approach.
This result thus shows the, at least formal, equivalence of these two RG approaches in the two dimensional case.
The only, relevant, difference is that the FRG approach furnishes an explicit representation for both the beta functions,
via equation (\ref{RG_3.1}), and for Zamolodchikov's metric, via equation (\ref{Zam}).

%%%%%%%%%%%%%%%%%
\section{Conclusion and Outlook}
%%%%%%%%%%%%%%%%%

In this paper we have clarified the connection between the functional
(or exact) renormalisation group (FRG) and the local renormalisation group (LRG) in the two dimensional case, showing
that these two major techniques used to obtain information on QFTs
arbitrarily far away from criticality are compatible and interconnected.

The proof of this connection was based on a careful analysis of the
form of the scale dependent Wess--Zumino action. We have seen in particular
that it is not only sufficient to identify the correct terms which
reproduce the conformal anomaly at the fixed point: it is also important
to constrain the form of the possible $\beta$--terms, which are generated
along the flow and die out at its endpoints.
These $\beta$--terms, called this way since they are proportional to beta functions,
can be mapped into corresponding terms found in the LRG. In the latter framework,
these terms simply arise as additional couplings due to the introduction
of spacetime dependent sources. They are then connected to beta functions
with the help of the Wess--Zumino consistency conditions. In this way, the consistency conditions
imposed by Weyl symmetry are able to constrain the possible forms
of the beta functions of the theory, and thus they provide information
on its RG flow. As we already remarked, the physical idea lying behind
this fact is that the sources used in the LRG act effectively like
running couplings, whose RG scale has become spacetime dependent.
This means that the scale transformations get promoted to Weyl transformations,
which automatically satisfy Wess--Zumino consistency conditions due to the structure of the Weyl
group.

However, our analysis has shown that the constraints on the form of
the running Wess--Zumino action are fairly general, and in fact they can be
used directly into the FRG to get the same results. In particular,
one can show in general that the beta function of $c$ must be quadratic
in the beta functions of the primary deformation couplings, a result
usually obtained with the LRG. If the two points of view are put together,
then the FRG gives a constructive way to compute the Zamolodchikov--Osborn
metric $\chi_{ij}$ for specific field contents.

One important remark should be made, and regards the proper form of
the $c$--function. In the LRG, the proper $c$--function is not simply
the running central charge, or anomaly coefficient  $\mathcal{C}_{k}$, but is instead
$\mathcal{C}_{k}+\omega_{i}\beta^{i}$. However, as we saw form the general
analysis in Section 3, this appears as the coefficient of the $\partial_{\mu}\tau\partial^{\mu}\tau$
term off--criticality, and at the FRG level it is quite indifferent
how we choose to represent this coefficient: all we really need is
to identify the proper monomial and then check its running. Now in
two dimensions, if we project the running on a flat background, that
choice is essentially unique, so we can pick that as our candidate
$c$--function. In higher dimension such as $d=4$ this choice will
in general be non--unique, so one must be more careful, but the same
remark applies.

Our analysis can naturally be extended to $d=4$ or higher dimensions,
with the proviso just made. This would be a particularly nice application
since little is known in this case about the behaviour of the flow
of the central charge far away from a conformal phase. We plan to
explore these issues in a future work.

\end{document}